\documentstyle[a4,12pt,epsf]{article}
\newcommand{\delfour}{{\Delta^{(4)}}}
\newcommand{\delsq}{\Delta^{(2)}}
\newcommand{\xv}{\vec{x}}

\newcommand{\mbare}{M^{0}_{c}}
\newcommand{\sigmav}{\mbox{\boldmath$\sigma$}}

\begin{document}

\title{$B_c$ Spectroscopy {}From Lattice QCD}

\author{
C.~T.~H.~Davies,$^a$
K.~Hornbostel,$^b$
G.~P.~Lepage,$^c$ \\
A.~J.~Lidsey,$^a$
J.~Shigemitsu,$^d$
J.~Sloan$^e$ \\[.4cm]
\small $^a$University of Glasgow, Glasgow, UK G12 8QQ, UKQCD collaboration.\\
\small $^b$Southern Methodist University, Dallas, TX 75275. \\
\small $^c$Newman Laboratory of Nuclear Studies, Cornell University,
Ithaca, NY 14853. \\
\small $^d$The Ohio State University, Columbus, OH 43210. \\
\small $^e$Florida State University, SCRI, Tallahassee, FL 32306. \\
}

\maketitle

\begin{abstract}
\noindent
We present first results for $B_c$ spectroscopy using Lattice
Non-Relativistic QCD (NRQCD). For the NRQCD action the
leading order spin-dependent and next to leading
order spin-independent interactions have been included with
tadpole-improved coefficients.
We use multi-exponential fits to multiple correlation
functions to extract ground and excited $S$ states
and give accurate values for the $S$ state hyperfine splitting
and the P state ($B^{**}_c$) fine structure, including the effects
of $^1P_1/^3P_1$ mixing.

\end{abstract}

\section{Introduction}

There is much current interest in the spectrum of mixed bound states of bottom
and charm quarks with the recent appearance of experimental
 candidates\cite{aleph}
Lattice QCD is the best method of calculating this spectrum from
first principles, although attention must be paid to the control
of systematic errors from a variety of sources.
We have recently given accurate results for bottomonium and charmonium
spectroscopy on the lattice \cite{ups,psi}, exploiting the non-relativistic
nature of these systems by using a non-relativistic effective theory,
NRQCD\cite{nrqcd}. This can be
systematically matched to full QCD, order by order
in $\alpha_s$ and $v^{2}/c^{2}$, where $v$ is the velocity of the
heavy quark in the bound state.
Here we present results for $B_c$ spectroscopy using the same techniques and
use our previous results to give good indications of
the size and direction
of systematic errors. We compare our results to recent
calculations in potential models \cite{eichten}.

\section{Lattice NRQCD}

The starting point of NRQCD is to expand the original QCD
Lagrangian in powers of $v^{2}$, the typical quark velocity in
a bound state.
For the $J/\Psi$ system $v^{2} \sim 0.3$ and for $\Upsilon$
$v^{2} \sim 0.1$. For $B_c$ $v^{2} \sim 0.15$ for the single
particle with mass equal to the reduced mass of the $b$, $c$ system,
but the kinetic energy will be shared unequally between the $b$ and
$c$. $v^{2} \sim 0.5$ for the $c$ quark so relativistic corrections will
be more important than for $c\overline{c}$ systems. NRQCD enables these
corrections to be added systematically.

The action used here is
the same one used in refs. \cite{ups} and \cite{psi}. Relativistic corrections
 ${\cal O} (M_Qv^{4})$ have been included for both quarks - errors are
then at the level of ${\cal O} (M_c v_c^6)$.
Other sources of systematic error are
discretisation errors and errors from
the absence of virtual quark loops
because we use
quenched configurations generated with the standard
plaquette action. Finite volume errors
should be negligible because of the relatively small size of heavy-heavy
 systems.

To calculate masses for $b\overline{c}$ bound states we define
$b$ and $c$ quark Green functions on the lattice.
The NRQCD Lagrangian
involves a simple difference equation in the temporal direction,
which allows the evolution of the quark Green function
as an initial value
problem.
We start on the first time slice with
\begin{eqnarray}
 G_1 &=&
  \left(1\!-\!\frac{aH_0}{2n}\right)^{n}
 U^\dagger_4
 \left(1\!-\!\frac{aH_0}{2n}\right)^{n} \, \delta_{\xv,0}
\end{eqnarray}
and then continue to evolve in the temporal direction using
\begin{eqnarray}
  G_{t+1} &=&
  \left(1\!-\!\frac{aH_0}{2n}\right)^{n}
 U^\dagger_4
 \left(1\!-\!\frac{aH_0}{2n}\right)^{n}\left(1\!-\!a\delta H\right) G_t
 \quad (t>1) .
\label{tevolve}
\end{eqnarray}
$H_0$ is the lowest order piece of the non-relativistic Hamiltonian, i.e.
the kinetic energy operator. On the lattice, this is
\begin{equation}
 H_0 = - {\delsq\over2\mbare}.
 \end{equation}
The higher order terms in the Hamiltonian that we have included are
\cite{cornell}
 \begin{eqnarray}
\label{deltaH}
\delta H
&=& - c_1 \frac{(\delsq)^2}{8(\mbare)^3}
            + c_2 \frac{ig}{8(\mbare)^2}\left({\bf \Delta}\cdot{\bf E} -
{\bf E}\cdot{\bf \Delta}\right) \nonumber \\
 & & - c_3 \frac{g}{8(\mbare)^2} \sigmav\cdot({\bf \Delta}\times{\bf E} -
{\bf E}\times{\bf \Delta})
 - c_4 \frac{g}{2\mbare}\,\sigmav\cdot{\bf B}  \nonumber \\
 & &  + c_5 \frac{a^2\delfour}{24\mbare}
     -  c_6 \frac{a(\delsq)^2}{16n(\mbare)^2} .
\end{eqnarray}
The first two terms in $\delta H$ are spin-independent
relativistic corrections and the next two are spin-dependent
terms which contribute the leading order pieces to
 the P and S spin splittings respectively. The last two terms
come from discretisation
corrections to the lattice Laplacian and the lattice time derivative.
${\bf \Delta}$ is the symmetric lattice derivative, $\delsq$
is the lattice form of the Laplacian and
 $\delfour$ is a lattice version of the continuum
operator $\sum D_i^4$. We use the standard traceless
cloverleaf operators for the
chromo-electric and magnetic fields, $\bf E$ and~$\bf B$.  The
parameter~$n$
is introduced to remove instabilities in the heavy quark
propagator caused by the highest momentum modes of the theory\cite{nrqcd}.
We require $n > 3/Ma$.
To calculate we must fix both the bare coupling and the bare
quark masses. We used 200 quenched gluon field configurations of size
$12^3 \times 24$
at $\beta = 6/g^2$ = 5.7 generously supplied by
the UKQCD collaboration \cite{thanks_ukqcd}, and fixed to Coulomb gauge,
using a Fourier accelerated steepest descents algorithm \cite{facc}.
 The bare masses of the $c$ and
$b$ quarks were chosen from fits to the kinetic mass of the
$\eta_c$ and the $\Upsilon$ \cite{psi, us_in_progress}.
At $\beta = 5.7$  we used a mass for the
$c$ quark in lattice units of 0.8 with $n = 4$, and a mass for the
$b$ quark of 3.15 with $n = 2$.

The coupling constants $c_{i}$ appearing in
equation (\ref{deltaH}) can be calculated by matching NRQCD to
full QCD \cite{cornell,colin}. At tree level all the coefficients are one.
The largest radiative corrections are believed to be
tadpole contributions which can be removed by redefining
the gluon fields $U$ :\cite{maclep}
\begin{eqnarray}
\label{US}
U_{\mu}(x) \rightarrow \frac{U_{\mu}(x)}{u_{0}}
\end{eqnarray}
with $u_{0}$ the fourth root of the plaquette (at $\beta$=5.7 we
use $u_0$ = 0.861). Since the cloverleaf
expression involves the evaluation of a plaquette this renormalization
will have a large effect on $E$ and $B$ fields and
thereby on spin-dependent splittings\cite{ups}.
With the dominant tadpole contributions thus removed, we
use the tree level values for the $c_{i}$'s.

Given the quark propagators in equation (\ref{tevolve})
it is relatively straightforward
to combine them appropriately to form meson propagators
with specific quantum numbers. This is outlined in \cite{ups}.
Here we must combine a $b$ quark propagator with a
$\overline{c}$ propagator, but the methods are identical.
We use `smearing functions' as sources for the $b$ quark
propagator and a delta function source for the $c$ quark.
Various different smearing functions are included to study
$p$ states and excited $s$ states. It is more efficient, from
equation (\ref{tevolve}), to have numerous $b$ quark propagators with a small
value of $n$ and only one $c$ quark propagator. It also
enabled us to simultaneously calculate the $b\overline{b}$ spectrum
on these lattices, and this will be reported separately \cite{us_in_progress}.
The radius of the smearing functions (taken as wavefunctions from
a $1/r$ potential) was kept fixed at
an optimal value for $b\overline{b}$ correlation functions. This
meant that the effective mass in the $B_c$ correlation
functions did not plateau
quite as early as it would have done with more
optimal smearing but it did not
significantly detract from the accuracy of our results.

Local meson operators are tabulated in \cite{ups}.
Using the notation $^{2S+1}L_{J}$, we have looked at meson
propagators for the following states: $^{1}S_{0}$, $^{3}S_{1}$, $^{1}P_{1}$,
$^{3}P_{0}$, $^{3}P_{1}$ and $^{3}P_{2}$ for both the
E and T representation.
Since $C$ is not a good quantum number for the $B_c$ system the
$^{1}P_1$ and $^{3}P_1$ mesons will mix and so in addition we
calculated the cross-correlation between these two.
For the
$s$ states, smearing functions both for the ground and first radially excited
state were used as well as a local $\delta$ function. {}From
this all possible combinations of smearing at the source with a
local sink were formed.
For the $p$ states only the ground state
smearing function was used at the source.
We calculated the dispersion relation for the $^{1}S_{0}$ ($B_c$)
by looking at the meson propagator
for small non-zero momenta.
Because of the relatively small size of these systems it is possible to
use more than one starting site for the mesons. We used 8 different
spatial origins and 2 different starting times to increase our statistics,
but we bin correlation functions over each configuration.

\section{\bf Simulation results}

For $b\overline{b}$ and $c\overline{c}$ the correlation functions are
explicitly real because of charge conjugation symmetry
- for $b\overline{c}$ they are not. We find, however,
that the imaginary parts of all our correlation functions are very small
and much smaller than the real parts. We fit simultaneous
multi-exponential fits to the real parts of several correlation functions
with different sources and local sinks. This is the `row' fit of
refs. \cite{ups,psi} :
\begin{equation}
G_{meson}(n_{sc},loc;t) = \sum_{k=1}^{N_{exp}}\; b(n_{sc},k)\,
e^{-E_k \cdot t } .
\label{mcorfit}
\end{equation}
$n_{sc}$ denotes the source smearing, $loc$ denotes a local sink.
It enables us to extract masses in lattice units for ground and
excited $s$ states. The $p$ states are much noisier and for
them we have used single exponential fits to the correlation
function with ground state smearing at both source and sink.
  For spin splittings it is
most accurate to perform a single exponential fit to a ratio of
correlation functions. We generate a bootstrap ensemble of ratios
(including both real and imaginary parts of the correlation functions)
and then fit to the real part.

To find masses for the eigenvectors of the $^1P_1$ and $^3P_1$ mixing
matrix we fit a 2-exponential form to the $2 \times 2$ matrix
formed from these correlation functions smeared at source and
sink. Figure 1 shows effective amplitude plots and fits for the
four correlation functions in the $2 \times 2$ matrix. The physical states are
called the $1^{+}$ and the $1^{+'}$. To calculate
the splitting between these two states and the mixing angle
with the $^1P_1/^3P_1$ basis accurately,
we did matrix fits to an ensemble
of matrices generated by bootstrap. This allows the correlation
between the states on a given configuration to be taken
into account. We checked these fits against those obtained
by diagonalising the matrix time-slice by time-slice and fitting
single exponential forms to the eigenvalues. There was good agreement
between the two methods.  We did not find it possible to extract
the masses of the $1^{+}$ and $1^{+'}$ purely from the $^1P_1$ and
$^3P_1$ correlation functions (i.e. the diagonal terms of the matrix)
in the absence of the cross-correlation terms. Although theoretically
possible this obviously requires very high statistics or a more
finely grained lattice in time.

\begin{figure}[t]
\epsfxsize = 12.0cm
\epsfbox[30 30 530 530]{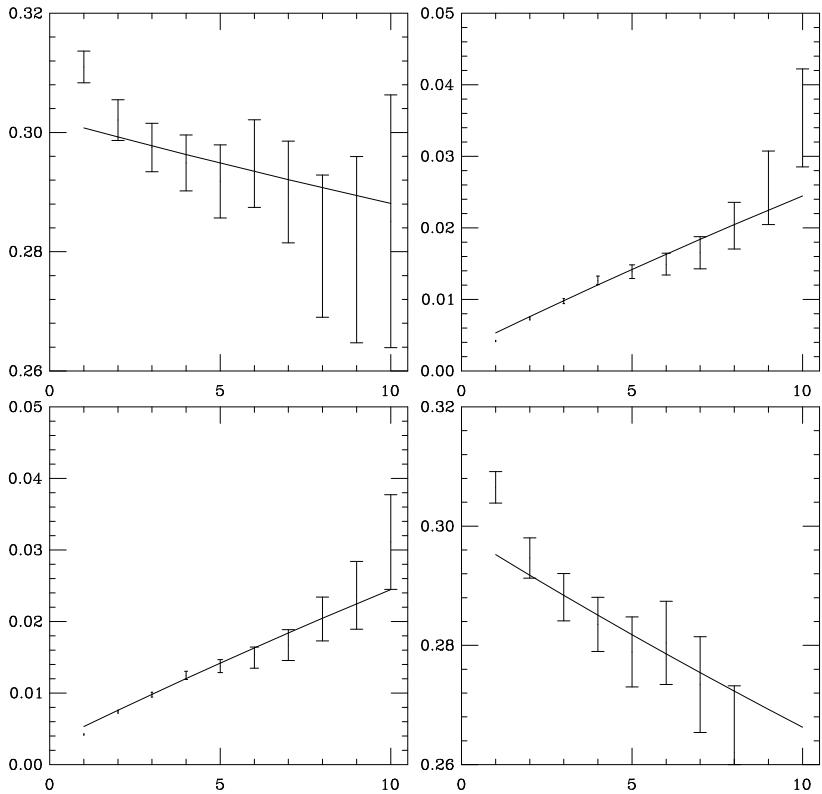}
\vspace{-0.2in}
\caption{ Effective amplitudes (obtained by multiplying the
correlation function by exp($M_{ground} t$)) for the four
correlation functions in the $2 \times 2$ $ ^1P_1 / ^3P_1$
mixing matrix. }
\label{effamp}
\end{figure}

Table 1 shows our results for masses and splittings in lattice units.
The errors quoted are statistical errors only.

\begin{table}
\begin{center}
\begin{tabular}{c|c}
& Simulation Results \\
\cline{1-2}
$1^{1}S_{0}$ & 0.6052(8) \\
$1^{3}S_{1}$ & 0.6353(9) \\
$2^{1}S_{0}$ & 1.12(6) \\
$2^{3}S_{1}$ & 1.14(6) \\
$1\ ^3P_2$ & 0.986(7) \\
$1\ ^3P_0$ & 0.944(7) \\
$1\ 1^{+}$ & 0.956(6) \\
$1\ 1^{+'}$ & 0.973(7) \\
\cline{1-2}
$^{3}S_{1} - ^{1}S_{0}$ & 0.0307(2) \\
$^{3}P_{2} - ^{3}P_{0}$ & 0.045(4) \\
$^{3}P_{2} - 1^{+}$ & 0.023(2) \\
$1^{+'} - 1^{+}$ & 0.0172(25) \\
$tan(\theta)$ & 0.66(3) \\
\cline{1-2}
$|\psi(0)|_{B_c}$ & 0.2118(4) \\
\cline{1-2}
\end{tabular}
\end{center}
\caption{Fitted dimensionless energies for $b\overline{c}$ states and
mixing angle for the $1^{+}$ state. $\theta$ is defined so that
$ | 1^{+} \rangle = sin(\theta) | ^1P_1 \rangle +
cos(\theta) | ^3P_1 \rangle$. The final entry is the wavefunction at the
origin for the $B_c$.}
\label{bcsimresults}
\end{table}

To convert the masses and splittings in lattice units to physical
results, we need a value for $a^{-1}$, the inverse lattice spacing.
Heavy-heavy systems provide a particularly good way of
fixing $a^{-1}$ because the spin-averaged 1P-1S splitting is
insensitive to the other parameter in the theory, i.e. the heavy
quark mass, and also to any systematic errors in spin-dependent terms.
Thus $a^{-1}$ can be fixed from a comparison of
either the $c\overline{c}$ or the $b\overline{b}$ spectrum to
experiment. Experimentally the spin averaged splitting is 452 MeV from
the $b\overline{b}$ spectrum and almost the same, 458 MeV,
 from the $c\overline{c}$ spectrum. Here the spin-averaged 1P mass is
taken as the average of the $\chi_b$ states, since only a preliminary
value for the mass of the $h_c$ is known and no $h_b$ has been seen \cite{pdg}.
Potential models also lead to the belief that the $^1P_1$ state should
lie approximately at the spin-average of the $^3P$ states, and so does not
need to be included.

Unfortunately, in the quenched approximation,
 the value obtained for $a^{-1}$ depends
on the experimental quantity you use to fix it, and in
particular, the momentum scales important to that quantity. This is because the
effective coupling constant does not run correctly between
different momentum scales. In general you expect that a quantity
which is more sensitive to large momenta will give a higher value
for $a^{-1}$. This is what we find on comparing the $b\overline{b}$ and
$c\overline{c}$ spectrum on quenched lattices. At $\beta$ = 5.7
$a^{-1}_{c\overline{c}}$ = 1.20(4) GeV \cite{psi} and $a^{-1}_{b\overline{b}}$
=
1.35(4) GeV \cite{us_in_progress}. These results
use the
spin-average of all $1P$ states \cite{ainv}
because we find that the $^1P_1$ does
not lie at the spin average of the $^3P$ states. This is likely to be, at
least partly, a
discretisation error \cite{psi}.

This discrepancy in values obtained for $a^{-1}$ from the quenched
approximation means that the only way to obtain results in good
agreement with experiment for a particular set of hadrons
is to fix $a^{-1}$ separately for that set
using one experimental result for a `typical hadron' from the set.
For $b\overline{c}$ systems there are no experimental results so
we must fix $a^{-1}$ on the basis of the numbers for $b\overline{b}$ and
$c\overline{c}$.
We assume that the spin-averaged 1P-1S splitting for $b\overline{c}$ will
have a value
between that for $b\overline{b}$ and
that for $c\overline{c}$ (which hardly differ, as above).
 We must now define the spin-averaged P state
from our results to include all 1P states since there is no reason
for either of the $1^{+}$ or the $1^{+'}$ to lie at the spin-average
of the other states.
This gives $a^{-1}$ = 1.32(4) GeV, as expected between the values from
$c\overline{c}$ and $b\overline{b}$. The error quoted is purely
statistical.

The difference in $a^{-1}$ between these different
systems leads to the effect that the quark mass in physical
units will differ between $b\overline{c}$ and either $b\overline{b}$
or $c\overline{c}$.
This problem has already been pointed out
for heavy-light systems \cite{lat94} and
means that again, in principle, one should fix the quark mass, using
experiment, separately
within each system in the quenched approximation.
 For the $B_c$ system, this is not possible, so
final meson masses will have an additional
uncertainty associated with this $a^{-1}$ effect. It will be important
particularly for spin splittings since these are strongly dependent
on the quark mass. We have fixed the bare quark masses in lattice
units from results for $b\overline{b}$ and $c\overline{c}$.
Another reasonable possibility is that we should readjust the
bare quark masses in $b\overline{c}$ so that they agree in
physical units with those of $b\overline{b}$ and $c\overline{c}$.
We have estimated what changes this might give where possible.

Using the value for $a^{-1}$ above, we can convert meson masses to physical
units. We still need to set a zero of energy since we have removed it
from our Hamiltonian. We do this by calculating $B_c$ correlation
functions at small but non-zero values of the meson momentum, {\bf P}, and
fitting to
\begin{equation} \label{dispers}
E_{\bf P} - E_{0} = \frac{{\bf P}^{2}}{2M_{kin}}.
\end{equation}
$M_{kin}$ is the kinetic (absolute) mass of the $B_c$ in lattice units.
We find a kinetic mass of 4.76(2), which, using
$a^{-1}$ as above, gives a mass in physical units of 6.28(20) GeV,
where the error is dominated by the uncertainty in $a^{-1}$.
Errors in the $b$ and $c$ quark masses discussed above actually cancel
in the kinetic mass of the $B_c$, since to match the bare physical
masses from $b\overline{b}$ and $c\overline{c}$ we in fact have to adjust
the $c$ mass down and the $b$ mass up by the same amount.

The difference between $E_0$ and $M_{kin}$ represents the shift of the zero
of energy. This can be calculated per quark in perturbation theory and
agreement with lattice results is very good for the $\Upsilon$ spectrum
\cite{ups}. For $c$ quarks the perturbative results are not very reliable
because the important scales for $\alpha_s$ are very low. However, we
can still check that the non-perturbative energy shift for the $B_c$ is
the average of those for $\Upsilon$ and $\Psi$. This should be the case
if no non-perturbative effects specific to the mesons appear, and
the results in Table 2 show that this is true. It is a requirement for
the non-relativistic theory to make sense and is apparently not obeyed
for heavy quark actions based on the Wilson action in current use\cite{sara}.

\begin{table}
\begin{center}
\begin{tabular}{c|lll}
 & $E_0$ & $M_{kin}$ & shift  \\
\cline{1-4}
$\Upsilon$  & 0.5030(5)  & 6.97(8) & 6.47(8)  \\
$\eta_c$  & 0.618(1) & 2.430(6) & 1.812(6) \\
$B_c$ & 0.6052(8) & 4.76(2) & 4.15(2) \\
\cline{1-4}
\end{tabular}
\end{center}
\caption{ Results for static and kinetic masses and the difference
between them for heavy-heavy mesons, using NRQCD at \protect $\beta$
=5.7. Values are in lattice units.
}
\label{shifts}
\end{table}

Table 3 gives results for the meson splittings with the
ground state ($B_c$) in GeV and the spectrum
is plotted in Figure 2 (using the $B_c$ mass of 6.28 GeV),
 with a comparison to results from a recent
potential model analysis \cite{eichten}.

\begin{table}
\begin{center}
\begin{tabular}{c|l}
 & Simulation Results [GeV]  \\
\cline{1-2}
$1{^3S}_{1} - B_c$  &  0.0405(3) \\
$2{^1S}_0 - B_c$ & 0.68(8) \\
$2{^3S}_1 - B_c$ & 0.71(8) \\
$1{^3P}_0 - B_c$ & 0.447(9) \\
$1\ 1^{+} - B_c$ & 0.463(8) \\
$1\ 1^{+'} - B_c$ & 0.485(8)  \\
$1{^3P}_2 - B_c$ & 0.503(8) \\
\cline{1-2}
\end{tabular}
\end{center}
\caption{NRQCD results for splittings of $b\overline{c}$ states
with $B_c$
for $a^{-1}$ = 1.32 GeV. Errors shown are statistical only.
}
\label{bcresults}
\end{table}

\section{Discussion}

Figure 2 shows the lattice spectrum compared to recent potential
model results \cite{eichten}. The lattice results have systematic
errors resulting from higher order relativistic corrections to the
action, discretisation effects and errors from using the quenched
approximation. We will attempt to quantify them below. The potential
model results have systematic errors also from various sources
including variations in the potential itself
 (compare, for example \cite{russians}). The lattice results
have the advantage that the errors there can, and will, be systematically
removed.

The main source of error for our lattice calculations is in higher order
relativistic corrections to
the charm quark propagator. The missing terms are at order $Mv^6$ and include
${\bf D}^6/16M^5_c$, relevant for spin-independent splittings.
A perturbative estimate of the size of this correction
would be an energy shift of $<p^6>/16M^5_c$.
A na\"{\i}ve evaluation of this
expectation value in a lattice potential model for $B_c$
gives 100 MeV. However, there are several terms at this order that should
be included and we would expect some cancellation between them. A
similar analysis of terms of order $Mv^4$, which are included here, shows
an almost complete cancellation between the ${\bf D}^4$ correction and the
Darwin term \cite{corrns}. Our comparison of lattice
results with and without these $Mv^4$ terms
shows agreement with this analysis \cite{ups}, giving us confidence that
we can estimate these corrections. The conclusion is then that terms of
order $Mv^6$ not included could shift our masses by 100 MeV, but by
cancellation it could be less than this.

The leading discretisation error is an $\cal{O}$$(a^2)$ error from using
gluon field configurations with the simple plaquette action. Again we can
estimate perturbatively the shift induced by this correction\cite{alpha}.
 The shift
is proportional to the square of the wave function at the origin,
 giving 15 MeV for the $B_c$, and zero for states with
no wave function at the origin such as $p$ states. The quark propagators
are already correct through $\cal{O}$$(a^2)$ and residual $a^4$ errors
should be smaller than this.

The errors from the quenched approximation should be between those for
$b\overline{b}$ and $c\overline{c}$ and can be estimated from the
differences we find in simulating those systems with and without
dynamical fermions. One source of error, that of uncertainties in the
quark masses, has been discussed above.

{}From Figure 2 the absolute mass of the $B_c$ which we obtain agrees well with
potential model estimates. Since it is close to the average of the $\eta_b$
and $\eta_c$ masses this is not surprising. The  spin-averaged 1P-1S
splitting also agrees by design, since we fixed it to be the same as
that for $b\overline{b}$ and $c\overline{c}$ and this would come out of
a potential model too. The possible 100 MeV (i.e. 25\%) systematic error in the
1P-1S splitting from higher order relativistic corrections should be
borne in mind. The corresponding systematic error for $b\overline{b}$ and
$c\overline{c}$ is 6 MeV and 40 MeV respectively.

The other spin-independent splitting is that between the 2S and 1S states.
All our quenched calculations \cite{psi,ups,us_in_progress} give a
ratio for the 2S-1S splitting over 1P-1S splitting which is too large.
This is expected  because the
1S state suffers a bigger shift downwards under quenching than the other
states.
We would therefore expect our 2S states to appear too high. For $b\overline{b}$
we find the 2S-1S/1P-1S ratio (using the $^3S_1$ for S)
to be 1.40(3) at $\beta$ = 6.0 \cite{lat95}
and 1.4(2) at $\beta$ = 5.7 \cite{us_in_progress}
compared to the experimental
value of 1.28 and for $c\overline{c}$ we get 1.4(2)
 \cite{psi} against the experimental
value of 1.38.
So, we would expect that unquenching would correct this
ratio for $B_c$ by roughly 10\%. Since we use the 1P-1S
splitting to fix $a^{-1}$, this means that our 2S-1S
splitting would be reduced by 10\%, i.e. our 2S level would fall by 70MeV.
This brings it rather closer to the potential model prediction, which
presumably already incorporates some effects from
dynamical fermions through the phenomenological form of the potential.
Large relativistic corrections for $B_c$ could distort this picture
somewhat.

The spin splittings are perhaps of more immediate interest and particularly
that between the $B_c^{*}$ and the $B_c$, since these states will be the
first to be studied experimentally in detail.
We find 40 MeV for this splitting in this simulation.
Again we expect a sizeable error from the quenched approximation. For
$b\overline{b}$ we have results both for dynamical flavors, $N_f$ = 0
and 2 and find that the hyperfine splitting changes by a factor
of 40 \% on unquenching. For $c\overline{c}$ the hyperfine splitting is
known and our quenched simulation gives a result 20 \% too low.
So we conclude that the splitting between the $B_c^{*}$ and $B_c$
might reasonably change to 55 MeV on unquenching. It will increase
by an additional 10\% to 60 MeV if the charm quark mass
in the $B_c$ is reduced so that it matches the bare value in
physical units for the $J/\Psi$. Both of these effects
again tend to improve the
agreement with the potential model (perturbative) value.

For the $p$ fine structure quenching effects may not be very large
because of the small wavefunction at the origin. More serious are
discretisation errors since the $p$ fine structure is the result
of a
balance between short and long range effects. {}From work
on $b\overline{b}$ \cite{ups,us_in_progress} and $c\overline{c}$ \cite{psi}
we find a tendency for lattice NRQCD calculations to underestimate
the overall size of the splittings $[M(^3P_2)-M(^3P_0)]$, a
larger effect for $b\overline{b}$ at $\beta$ = 5.7 than
for $c\overline{c}$. We believe that this is because the short-
range components of the spin-orbit and tensor forces are
underestimated at this lattice spacing. If so, the
$^3P_2 - ^3P_0$ splitting will also be underestimated here,
from the same effect. The result for the splitting would
additionally increase if the charm quark mass were reduced.
Our value for this splitting is 60(5) MeV, already
larger than some potential model results. It seems likely to
us that the experimental value will exceed 60 MeV.

Another result from our $b\overline{b}$ and $c\overline{c}$
work was the ratio of fine structure splittings $[M(\chi_2) -
M(\chi_1)] / [M(\chi_1) - M(\chi_0)]$. This exceeded the experimental
values and was another indicator that the long-range
spin-orbit force has undue dominance at this lattice spacing.
For the perturbative potential model of ref. \cite{eichten}
this ratio comes out below experiment for charmonium (although
not for bottomonium),
indicating that perhaps the long-range spin-orbit term is
not large enough.

For $B_c$ the $\chi_1$ state mixes with the $^1P_1$.
Our results
show such a large amount of mixing that the physical states
$1^{+}$ and $1^{+'}$ are very close to the $jj$ coupled states
expected in the $M_b \rightarrow \infty$ limit. For these states we
couple $\vec{L} + \vec{s_c} = \vec{J_c}$ and $\vec{J_c} +
\vec{s_b} = \vec{J}$. The $J_c$ = 1/2 state has a mixing
angle $\theta$ with the $LS$ coupled basis of tan($\theta$) =
1/$\sqrt(2)$ = 0.71.  In the $M_b \rightarrow \infty$ limit
the $1^{+}$ state becomes the
$J_c$ = 1/2 state, degenerate with the $\chi_0$,
and the $1^{+'}$, the $J_c$ = 3/2 state, degenerate with the $\chi_2$.
Our result of tan($\theta$) = 0.66 is quite different to the (perturbative)
potential model \cite{eichten} in which the mixing (viewed
from the $LS$ coupled basis) is very small.

The true result probably lies somewhere in between. In the
$LS$ basis and in potential model language, the off-diagonal
terms of the mixing matrix are provided purely by the
long-range spin-orbit potential. The short-range spin-orbit
and tensor terms contribute only to the diagonal pieces
(as does the spin-spin term, although we expect it to vanish
for $p$ states). For our calculation then the off-diagonal
terms are too large compared to the on-diagonal, and we
overestimate the mixing. For the perturbative potential models,
the opposite may be true.
The mixing is clearly very sensitive to these effects.

We can also estimate the decay constant $f_{B_c}$ from our result
for the wavefunction at the origin. Using the standard formula $f^2 =
12 |\psi(0)|^2 / M $, we obtain $f_{B_c}$ = 440(20) MeV, where the
error quoted is dominated by the uncertainty in $a^{-1}$.
This result is only valid to leading order in the inverse quark masses
and $\alpha_s$.
Relativistic $1/M^2_c$ corrections should be applied
as well as a perturbative renormalisation factor
before a useful continuum value can be quoted \cite{kim}.

% B_C compared to Eichten and Quigg potential model
%\Large
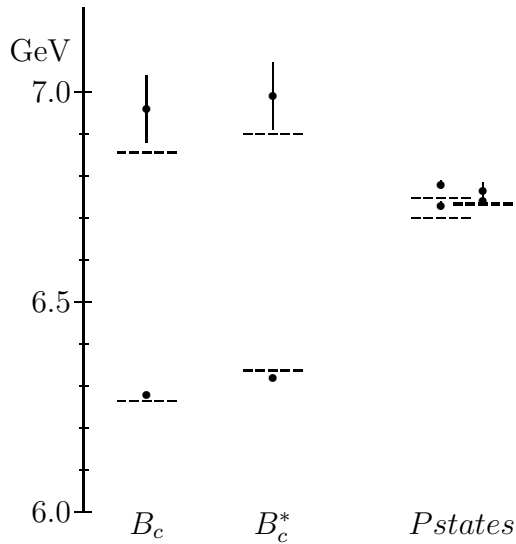
\begin{figure}
\begin{center}
\setlength{\unitlength}{.022in}
\begin{picture}(130,120)(10,580)
% axis
\put(15,600){\line(0,1){120}}
\multiput(13,600)(0,50){3}{\line(1,0){4}}
\multiput(14,600)(0,10){11}{\line(1,0){2}}
\put(12,600){\makebox(0,0)[r]{6.0}}
\put(12,650){\makebox(0,0)[r]{6.5}}
\put(12,700){\makebox(0,0)[r]{7.0}}
\put(12,710){\makebox(0,0)[r]{GeV}}

     \put(30,600){\makebox(0,0)[t]{$B_c$}}
     \put(30,628){\circle*{2}}
     \multiput(23,626.4)(3,0){5}{\line(1,0){2}}

     \put(30,696){\circle*{2}}
     \put(30,696){\line(0,1){8}}
     \put(30,696){\line(0,-1){8}}
     \multiput(23,685.6)(3,0){5}{\line(1,0){2}}

     \put(60,600){\makebox(0,0)[t]{$B^{*}_c$}}
     \put(60,632){\circle*{2}}
     \multiput(53,633.7)(3,0){5}{\line(1,0){2}}

     \put(60,699){\circle*{2}}
     \put(60,699){\line(0,1){8}}
     \put(60,699){\line(0,-1){8}}
     \multiput(53,689.9)(3,0){5}{\line(1,0){2}}

     \put(105,600){\makebox(0,0)[t]{$P states$}}
     \put(100,673){\circle*{2}}
     \put(100,673){\line(0,1){1}}
     \put(100,673){\line(0,-1){1}}
     \multiput(93,674.7)(3,0){5}{\line(1,0){2}}

     \put(100,678){\circle*{2}}
     \put(100,678){\line(0,1){1}}
     \put(100,678){\line(0,-1){1}}
     \multiput(93,670.0)(3,0){5}{\line(1,0){2}}

     \put(110,674){\circle*{2}}
     \put(110,674){\line(0,1){1}}
     \put(110,674){\line(0,-1){1}}
     \put(110,676.5){\circle*{2}}
     \put(110,676.5){\line(0,1){2}}
     \put(110,676.5){\line(0,-1){2}}
     \multiput(103,673.0)(3,0){5}{\line(1,0){2}}
     \multiput(103,673.6)(3,0){5}{\line(1,0){2}}

%     \put(165,335){\circle*{2}}
%\multiput(165,333)(0,3){2}{\line(0,1){1}}
%     \put(155,324){\makebox(0,0)[l]{expected}}
%     \put(155,317){\makebox(0,0)[l]{systematic}}
%     \put(155,310){\makebox(0,0)[l]{error}}

\end{picture}
\end{center}
\caption{NRQCD simulation results for the spectrum of the
$B_c$ system
using an inverse lattice spacing of 1.32~GeV.
Error bars are shown where visible and only indicate statistical
uncertainties. Dashed lines show results from a recent
potential model calculation \protect \cite{eichten}}
\label{fig:bc}
\end{figure}

\section{Conclusions}

This represents a first calculation of the $b\overline{c}$ spectrum
on the lattice.
 We use NRQCD and include the leading relativistic and discretisation
corrections with tadpole-improved coefficients.  We give a spectrum
including radially excited $s$ states as well as $p$ fine
structure taking account of mixing between the $J$ = 1 states.
Agreement with potential model results is surprisingly good,
given that they include, at least explicitly, no relativistic
corrections and the velocity of the charm quark within a
$B_c$ is actually higher than for charmonium.
Analysis of the systematic errors in the lattice calculation
tends to improve the agreement with potential models
{\it except} for the overall
scale of the $p$ fine structure, $M(^3P_2) - M(^3P_0)$. We
believe this is underestimated at present. Our result for
the mixed $J=1$ states show strong mixing in the $LS$ coupled basis
so that physical states are close to the
$jj$ coupled limit. We believe that the improvement
of systematic errors from discretisation will tend to reduce this mixing.
Future calculations of the spectrum will work at smaller
lattice spacings with relativistic charm quarks.

\vspace{4mm}

{\bf Acknowledgements}
This calculation was performed at the Atlas Centre under grant GR/J18927 from
the UK PPARC and at NERSC. AJL is also grateful to PPARC for a studentship,
CTHD for support under grant GR/J21231 and JSl for a Visiting Fellowship to
Glasgow
while this work was being completed.
This work was supported in part by grants from the U.S.Department of
Energy (DE-FC05-85ER250000, DE-FG05-92ER40742, DE-FG02-91ER40690), and
the National Science Foundation.
We thank UKQCD for
making their configurations available to us, and in particular David
Henty who helped us to read them.  We thank Ian Knowles for useful
discussions.


\begin{thebibliography}{8}
\bibitem{ainv} In refs. 1 and 2 above the quoted value for $a^{-1}$
is that using the $^1P_1$ mass for the spin-averaged 1P state, but
the other masses are also given. The differences in $a^{-1}$ are within
the statistical error.
\end{thebibliography}
\end{document}